\documentclass[9pt, conference]{IEEEtran}
\IEEEoverridecommandlockouts

\usepackage{cite}
\usepackage{amsmath,amssymb,amsfonts}
\usepackage{algorithmic}
\usepackage{graphicx}
\usepackage{textcomp}
\usepackage{xcolor}

\usepackage{graphicx}
\usepackage{dsfont,microtype,xcolor}
\usepackage{musicography}
\usepackage{xspace}
\usepackage{hyperref}

\definecolor{darkviolet}{rgb}{0.75,0,0.75}
\definecolor{forestgreen}{rgb}{0.18,0.55,0.34}
\def\BibTeX{{\rm B\kern-.05em{\sc i\kern-.025em b}\kern-.08em
    T\kern-.1667em\lower.7ex\hbox{E}\kern-.125emX}}

\newcommand\ourmethod{\xspace{S-KEY}\xspace}

\title{S-KEY: Self-Supervised Learning \\ of Major and Minor Keys from Audio}
\author{\IEEEauthorblockN{Yuexuan Kong$^{1,2}$, Gabriel Meseguer-Brocal$^1$, Vincent Lostanlen$^2$, Mathieu Lagrange$^2$, Romain Hennequin$^1$}
\IEEEauthorblockA{$^1$\textit{Deezer Research} \\
Paris, France\\
$^2$\textit{Nantes Université, Centrale Nantes, CNRS, LS2N, UMR 6004}\\
F-44000 Nantes, France
}
}
\begin{document}
\maketitle

\begin{abstract}
STONE, the current method in self-supervised learning for tonality estimation in music signals, cannot distinguish relative keys, such as C major versus A minor.
In this article, we extend the neural network architecture and learning objective of STONE to perform self-supervised learning of major and minor keys (\ourmethod).
Our main contribution is an auxiliary pretext task to STONE, formulated using transposition-invariant chroma features as a source of pseudo-labels.
\ourmethod{} matches the supervised state of the art in tonality estimation on FMAKv2 and GTZAN datasets while requiring no human annotation and having the same parameter budget as STONE.
We build upon this result and expand the training set of \ourmethod{} to a million songs, thus showing the potential of large-scale self-supervised learning in music information retrieval.
\end{abstract}

\section{Introduction}
Variations in tonality tend to elicit sensations of surprise among music listeners \cite{parncutt2024psychoacoustic}.
Characterizing these variations is a long-standing topic in music information retrieval (MIR), with MIREX serving as a standard evaluation framework in the case of Western tonality \cite{downie2010music}.
Yet, despite the interest in deep convolutional networks (convnets) in MIR \cite{humphrey2013feature}, they depend on a collection of expert annotations for supervised learning.
This is at odds with so-called \emph{implicit learning} in humans: explicit understanding of erudite concepts of music theory is not necessary to perceive harmonic contrast.
Hence, we question the need for supervision in machine learning for tonality estimation.

An alternative paradigm, known as self-supervised learning (SSL), has found promising applications into MIR \cite{liu2022audio}.
The gist of SSL is to formulate a \emph{pretext task}; i.e., one in which the correct answer may be inexpensively obtained from audio data.
While some SSL systems have general-purpose pretext tasks and require supervised fine-tuning \cite{liu2024audioldm,niizumi2021byol,spijkervet2021contrastive,li2024mert}, others are tailored for specific downstream tasks: e.g., the estimation of pitch \cite{gfeller2020spice,riou2023pesto}, tempo \cite{quinton2022equivariant,gagnere2024adapting}, beat \cite{desblancs2023zero}, drumming patterns \cite{choi2019deep}, and structure \cite{buisson2022learning}.

Very recently, a pretext task has been proposed for tonality estimation, as part of two SSL models: STONE, a key signature estimator, and its variant 24-STONE, the only existing self supervised key signature and mode estimator \cite{kong2024stone}.
However, STONE is incomplete in the sense that it is insensitive to modulations within a given key signature: for example, STONE may distinguish C major from A major or from C minor, but not from A minor. On the other hand, 24-STONE, as a first proposition toward self-supervised key signature and mode estimator, underperforms by 15\% when compared to models incorporating supervision.
The issue of coming up with an SSL technique which could classify key signatures as well as major and minor modes that can achieve comparable performance as supervised models remains as an open problem.

In this article, we present \ourmethod, the first SSL model that learns to represent both the distinction between key signatures and modes. Given that major and minor modes are the two most representative modes in western music, in this paper, we limit mode classification only to major and minor modes, which is often the case in literature\cite{korzeniowski2018genre}.
The main idea behind \ourmethod{} is to form pseudo-labels for the mode classification by comparing the chroma features which correspond to the root notes of the relative major and minor scales.
To identify these root notes, we rely on self-supervised knowledge about key signatures, as obtained via a STONE-like pretext task.
The originality of \ourmethod{} is to re-inject this knowledge into the formulation of a finer-grained task.
For simplicity and efficiency, our convnet optimizes both tasks at once, via a structured output for 24-class classification: 12 key signatures and two modes.

Our main finding is that \ourmethod{} achieves a MIREX score\cite{downie2010music} of 72.1\% on the FMAKv2 dataset, outperforming the self-supervised state of the art (SOTA) of 57.9\% held by 24-STONE with the same number of parameters and training samples (60k songs).
Scaling up SSL to 1M songs brings the MIREX score of \ourmethod{} up to 73.2\%, on par with the \emph{supervised} SOTA (73.1\%) of \cite{korzeniowski2018genre}.
We expand our MIREX-compliant benchmark to three other datasets: GTZAN, GiantSteps, and Schubert Winterreise Dataset (SWD).
Although key classification remains challenging for certain genres (\emph{e.g.}, blues, jazz, and hip-hop), \ourmethod{} is the first SSL method which matches or outperforms supervised deep learning for this task with no need for supervision.

\section{Methods}
Our proposed method builds on previous publication \cite{kong2024stone} whose key components are briefly presented in \ref{sub:chromanet} and \ref{sub:cpsd}. From \ref{sub:pseudo} to \ref{sub:all-together}, we introduce novel contributions of \ourmethod{} which replace the necessity of supervision in 24-STONE by self-supervision.

\subsection{Structured prediction with ChromaNet}
\label{sub:chromanet}

ChromaNet is defined as the combination of audio pre-processing, the 2-D convolutional neural network and the octave pooling.

For each song, we extract two disjoint time segments, denoted by $\mathrm{A}$ and $\mathrm{B}$.
We compute their constant-$Q$ transforms (CQT) with $Q=12$ bins per octave and center frequencies ranging between $27.5$ Hz and $8.37$ kHz (99 bins).
We denote the CQT of segment A by $\boldsymbol{x}_{\mathrm{A}}$ and idem for $\boldsymbol{x}_{\mathrm{B}}$, which are assumed to have the same key.

To perform artificial pitch transposition, we crop CQT rows in $\boldsymbol{x}_{\mathrm{A}}$ to simulate a pitch transposition by $c$ semitones for $0 \leq c\leq15$: $T_{c}\boldsymbol{x}_{\mathrm{A}}[p, t] = \boldsymbol{x}_{\mathrm{A}}[p-c, t]$ for each $c \leq p < QJ$ where $J=7$ octaves.
All CQTs after cropping result in $QJ=84$ bins in total.
$T_{0}\boldsymbol{x}_{\mathrm{A}}$ and $T_{k}\boldsymbol{x}_{\mathrm{A}}$ are assumed to have a pitch difference of $k$ semitones.

We define a 2-D fully convnet $f_{\boldsymbol{\theta}}$ with trainable parameters $\boldsymbol{\theta}$, operating on $T_{c}\boldsymbol{x}_{\mathrm{A}}$ and $T_{c}\boldsymbol{x}_{\mathrm{B}}$ with $M=2$ output channels and no pooling over the frequency dimension.
Over each channel, we apply average pooling on the time dimension and batch normalization.

The matrix of learnable activations $\boldsymbol{f}_{\boldsymbol{\theta}}(T_{c}\boldsymbol{x}_{\mathrm{A}})$ has $QJ=84$ rows and $M=2$ columns.
We sum this matrix across octaves, i.e., across rows by $Q$ semitones apart, and apply a softmax transformation over all $QM=24$ entries.

\begin{figure}
    \centering
    \includegraphics[width=\linewidth]{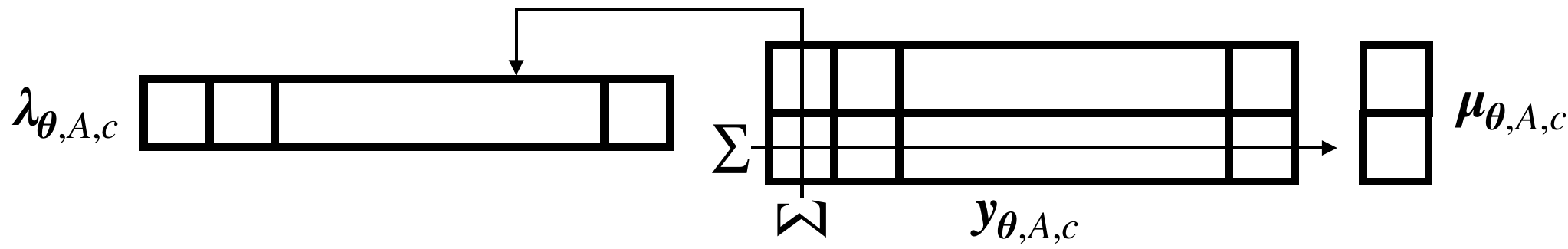}
    \caption{Structured prediction: Summing $\boldsymbol{y}_{\boldsymbol{\theta}, \mathrm{A}, \mathrm{c}}$ over rows produces a pitch-equivariant component $\boldsymbol{\lambda}_{\boldsymbol{\theta}, \mathrm{A}, \mathrm{c}}$, summing $\boldsymbol{y}_{\boldsymbol{\theta}, \mathrm{A}, \mathrm{c}}$ per columns produces a pitch-invariant component $\boldsymbol{\mu}_{\boldsymbol{\theta}, \mathrm{A}, \mathrm{c}}$. Rows and columns are reversed in the figure compared to the main text due to space limitation for the figure.}
    \label{fig:summing}
\end{figure}

This yields a matrix $\boldsymbol{y}_{\boldsymbol{\theta},\mathrm{A}, \mathrm{c}}$ with $Q=12$ rows and $M=2$ columns
whose entries are nonnegative and sum to one.
We sum the columns of  $\boldsymbol{y}_{\boldsymbol{\theta},\mathrm{A}, \mathrm{c}}$, yielding  $\boldsymbol{\lambda}_{\boldsymbol{\theta},\mathrm{A}, \mathrm{c}}[q] = \sum_{m=0}^{M-1}\boldsymbol{y}_{\boldsymbol{\theta},\mathrm{A}, \mathrm{c}}[q, m]$ a vector with $Q$ nonnegative entries summing to one.
Likewise over rows: $\boldsymbol{\mu}_{\boldsymbol{\theta},\mathrm{A}, \mathrm{c}}[m] = \sum_{q=0}^{Q-1}\boldsymbol{y}_{\boldsymbol{\theta},\mathrm{A}, \mathrm{c}}[q,m]$, a vector with $M$ nonnegative entries summing to one.
This is a kind of structured prediction: the learned representation $\boldsymbol{y}_{\boldsymbol{\theta},\mathrm{A}, \mathrm{c}}$ has a pitch-equivariant component $\boldsymbol{\lambda}_{\boldsymbol{\theta},\mathrm{A}, \mathrm{c}}$ and a pitch-invariant component $\boldsymbol{\mu}_{\boldsymbol{\theta},\mathrm{A}, \mathrm{c}}$, as shown in Figure \ref{fig:summing}.
Idem for $\boldsymbol{y}_{\boldsymbol{\theta},\mathrm{B}, \mathrm{c}}$, $\boldsymbol{\lambda}_{\boldsymbol{\theta},\mathrm{B}, \mathrm{c}}$, and $\boldsymbol{\mu}_{\boldsymbol{\theta},\mathrm{B}, \mathrm{c}}$.

\subsection{Cross-power spectral density (CPSD)}
\label{sub:cpsd}
The cross-power spectral density (CPSD) of $\boldsymbol{\lambda}_{\boldsymbol{\theta}, \mathrm{A}, \mathrm{c}}$ and $\boldsymbol{\lambda}_{\boldsymbol{\theta}, \mathrm{B}, \mathrm{c}}$ is the product $\widehat{\boldsymbol{\lambda}}_{\boldsymbol{\theta}, \mathrm{A}, \mathrm{c}}[\omega]\widehat{\boldsymbol{\lambda}}_{\boldsymbol{\theta}, \mathrm{B}, \mathrm{c}}^{\ast}[\omega]$, where the hat denotes a discrete Fourier transform (DFT), the asterisk denotes a complex conjugation, and the discrete frequency variable $\omega$ is coprime with 12. We set $\omega=7$ so that the phase of the CPSD coefficient denotes a key modulation over the circle of fifths (CoF)---see \cite{kong2024stone} for details.

Intuitively, while $\boldsymbol{\lambda}_{\boldsymbol{\theta}, \mathrm{A}, \mathrm{c}}$ is a one-hot encoding, $\widehat{\boldsymbol{\lambda}}_{\boldsymbol{\theta}, \mathrm{A}, \mathrm{c}}$ is a complex number of magnitude 1 on the CoF.
Given an integer $k$, the CPSD of $\boldsymbol{\lambda}_{\boldsymbol{\theta}, \mathrm{A}, \mathrm{c}}$ and $\boldsymbol{\lambda}_{\boldsymbol{\theta}, \mathrm{A}, \mathrm{c+k}}$ is the difference of phases corresponding to a pitch modulation of $k$ semitones on the CoF.

We define a CPSD-based function $\mathcal{D}_{\boldsymbol{\theta},c,k}$ which is equal to zero if and only if the vectors $\boldsymbol{\lambda}_{\boldsymbol{\theta},\mathrm{A}, \mathrm{c}}$ and $\boldsymbol{\lambda}_{\boldsymbol{\theta},\mathrm{B}, \mathrm{c+k}}$ contain a single nonzero coefficient and are equal up to circular shift by $k$:
\begin{equation}
\mathcal{D}_{\boldsymbol{\theta},c,k}(\boldsymbol{x}_{\mathrm{A}}, \boldsymbol{x}_{\mathrm{B}}) = \dfrac{1}{2} \left \vert
e^{- 2\pi\mathrm{i}\omega k/Q} - \widehat{\boldsymbol{\lambda}}_{\boldsymbol{\theta}, \mathrm{A}, \mathrm{c}}[\omega]\widehat{\boldsymbol{\lambda}}_{\boldsymbol{\theta}, \mathrm{B}, \mathrm{c+k}}^{\ast}[\omega]
\right \vert^2.
\label{eq:ddcf}
\end{equation}
For any integer $k$ and pair $\boldsymbol{x}=(\boldsymbol{x}_{\mathrm{A}},\boldsymbol{x}_{\mathrm{B}})$, $\mathcal{D}_{\boldsymbol{\theta},c,k}$ is differentiable with respect to ChromaNet weights $\boldsymbol{\theta}$.
Hence, we define a CPSD-based loss function\footnote{In this paper, we use the vertical bar notation to clearly separate neural network parameters on the left versus data and random values on the right.} which is parametrized by $c$ and $k$:
\begin{align}
\mathcal{L}_{\mathrm{CPSD}}(\boldsymbol{\theta}\,\vert\,\boldsymbol{x},c,k) &=
\mathcal{D}_{\boldsymbol{\theta},c,0}(\boldsymbol{x}_{\mathrm{A}}, \boldsymbol{x}_{\mathrm{B}}) \nonumber \\
&+ \mathcal{D}_{\boldsymbol{\theta},c,k}(\boldsymbol{x}_{\mathrm{A}}, \boldsymbol{x}_{\mathrm{A}}) \nonumber \\
&+ \mathcal{D}_{\boldsymbol{\theta},c,k}(\boldsymbol{x}_{\mathrm{B}}, \boldsymbol{x}_{\mathrm{A}}).
\label{eq:cpsd}
\end{align}
In Equation \eqref{eq:cpsd}, the first term encourages the model $\boldsymbol{f}_{\boldsymbol{\theta}}$ to be invariant to the permutation of $\boldsymbol{x}_{\mathrm{A}}$ and $\boldsymbol{x}_{\mathrm{B}}$, while the second and third term encourage it to be equivariant to the pitch interval $k$. As \cite{kong2024stone} points out, all three terms are indispensable for an efficient optimization of the model without collapsing into a uniform or constant distribution.
\subsection{Pseudo-labeling of mode}
\label{sub:pseudo}
STONE has shown that training a ChromaNet to minimize $\mathcal{L}_{\mathrm{CPSD}}$ produces a pitch-equivariant representation which is a sparse nonnegative vector in dimension $Q$.
We elaborate on this prior work to build a self-supervised approximate predictor of key signature, based on the pitch-equivariant component $\boldsymbol{\lambda}_{\boldsymbol{\theta}}$ for both segments A and B:
\begin{equation}
q_{\max}\left(\boldsymbol{\theta}\,\vert\,\boldsymbol{x}\right)=\arg\max_{0\leq q < Q}\left(\boldsymbol{\lambda}_{\boldsymbol{\theta}, \mathrm{A}, \mathrm{c}}[q] + \boldsymbol{\lambda}_{\boldsymbol{\theta}, \mathrm{B}, \mathrm{c}}[q]\right).
\end{equation}
Our postulate is that, if $\mathcal{L}_{\mathrm{CPSD}}(\boldsymbol{\theta})$ is low and $\boldsymbol{x}$ is in a major key, $q_{\max}\left(\boldsymbol{\theta}\,\vert\,\boldsymbol{x}\right)$ on the CQT scale corresponds to its root pitch class.


We compute a pitch class profile (PCP) for $\boldsymbol{x}$ by averaging its CQT across octaves, along time, and across segments A and B:
\begin{equation}
\boldsymbol{u}(\boldsymbol{x})[q] = \dfrac{1}{2} \sum_{j=0}^{J-1} \sum_{t=0}^{\tau-1}\big(\boldsymbol{x}_{\mathrm{A}}[Qj+q,t] + \boldsymbol{x}_{\mathrm{B}}[Qj+q,t]\big)
\end{equation}
Without side information nor learning, $\boldsymbol{u}(\boldsymbol{x})$ would be a poor predictor of tonality, as it erases spectrotemporal dynamics in $\boldsymbol{x}$.
However, when the key signature is known (e.g., no \musFlat{} nor \musSharp{}), comparing the CQT energy of the root note of the major key (e.g., C) with that of the relative minor key (e.g., A) can achieve an accuracy of 79.4\% in correctly determining the mode.
Our main idea for this paper is to use the key signature predictor $q_{\max}(\boldsymbol{\theta})$ as side information to improve pretext task design based on $\boldsymbol{u}(\boldsymbol{x})$.

We look up the entry 
$u_{\mathrm{maj}}(\boldsymbol{\theta}\,\vert\,\boldsymbol{x},c) = \boldsymbol{u}(T_{c}\boldsymbol{x})[q_{\max}(\boldsymbol{\theta}\,\vert\,\boldsymbol{x})]$,
where $T_{c}\boldsymbol{x}$ is a shorthand for $(T_{c}\boldsymbol{x}_{\mathrm{A}}, T_{c}\boldsymbol{x}_{\mathrm{B}})$.
Its value may be interpreted as the acoustical energy at the root pitch class under the assumption that the song is in a major key.
Conversely, we look up $u_{\mathrm{min}}(\boldsymbol{\theta}\,\vert\,\boldsymbol{x},c) = \boldsymbol{u}(T_{c}\boldsymbol{x})[(q_{\max}(\boldsymbol{\theta}\,\vert\,\boldsymbol{x})-3)\,\mathrm{mod}\,Q]$, i.e., idem under the assumption that the song is in a minor key.
Since $Q=12$, the number $3$ in the definition of $u_{\mathrm{min}}$ corresponds to a minor third, i.e., the interval between roots of relative keys.
We define a pseudo-label $\boldsymbol{\nu}$ for SSL of mode according to a simple logical rule:
\begin{equation}
\boldsymbol{\nu}(\boldsymbol{\theta}\,\vert\,\boldsymbol{x},c) =
\begin{cases}
[1,0] & \text{if $(u_{\mathrm{maj}}(\boldsymbol{\theta}\,\vert\,\boldsymbol{x},c) > u_{\mathrm{min}}(\boldsymbol{\theta}\,\vert\,\boldsymbol{x},c))$}\\
[0,1] & \text{otherwise.}
\end{cases}
\label{eq:nu}
\end{equation}

\subsection{Binary cross-entropy (BCE) with pseudo-labels}
Given $\boldsymbol{\nu}(\boldsymbol{\theta}\,\vert\,\boldsymbol{x},c)$ and $k$, we define a novel loss function:
\begin{align}
\mathcal{L}_{\mathrm{S-KEY}}(\boldsymbol{\theta}\,\vert\,\boldsymbol{x},c,k) &= \mathrm{BCE}(\boldsymbol{\nu}(\boldsymbol{\theta}\,\vert\,\boldsymbol{x},c),\boldsymbol{\mu}_{\boldsymbol{\theta},\mathrm{A}, \mathrm{c}})
\nonumber \\
&+ \mathrm{BCE}(\boldsymbol{\nu}(\boldsymbol{\theta}\,\vert\,\boldsymbol{x},c), \boldsymbol{\mu}_{\boldsymbol{\theta},\mathrm{B}, \mathrm{c}})
\nonumber \\
&+ \mathrm{BCE}(\boldsymbol{\nu}(\boldsymbol{\theta}\,\vert\,\boldsymbol{x},c), \boldsymbol{\mu}_{\boldsymbol{\theta},\mathrm{A}, \mathrm{c+k}})
\label{eq:bce-loss}
\end{align}
where $\mathrm{BCE}(\boldsymbol{\nu}, \boldsymbol{\mu}) = - \boldsymbol{\nu}[0] \log \boldsymbol{\mu}[0]
- \boldsymbol{\nu}[1] \log \boldsymbol{\mu}[1]$ denotes binary cross-entropy.
Intuitively, $\mathcal{L}_{\mathrm{S-KEY}}$ is low if and only if the structured predictions $\boldsymbol{f}_{\boldsymbol{\theta}}(T_{c}\boldsymbol{x}_{\mathrm{A}})$, $\boldsymbol{f}_{\boldsymbol{\theta}}(T_{c}\boldsymbol{x}_{\mathrm{B}})$, and $\boldsymbol{f}_{\boldsymbol{\theta}}(T_{c+k}\boldsymbol{x}_{\mathrm{A}})$ have large coefficients in the column corresponding to the pseudo-label $\boldsymbol{\nu}(\boldsymbol{\theta}\,\vert\,\boldsymbol{x},c)$.

Crucially, the equation above is different from the definition of $\mathcal{L}_{\mathrm{BCE}}$ in 24-STONE \cite[Equation 16]{kong2024stone}, which only involves pairwise BCE's between ChromaNet activations $\boldsymbol{\mu}_{\boldsymbol{\theta}}$.

While STONE is symmetric across columns, \ourmethod{} breaks this asymmetry via the pseudo-labeling function $\boldsymbol{\nu}$, making it less susceptible to model collapse. $\boldsymbol{\nu}$ replaced the indispensable supervision for 24-STONE to match the performance of supervised models.



\subsection{Loss over batch-wise average of mode predictions}

SSL training with $\mathcal{L}_{\mathrm{S-KEY}}$ faces a ``cold start'' problem in the sense that the pseudo-labeling function $\boldsymbol{\nu}$ is itself parametrized by the pitch equivariant component $\boldsymbol{\lambda}_{\boldsymbol{\mathrm{\theta}}}$, therefore ChromaNet weights $\boldsymbol{\theta}$.
During informal experiments, we have observed that penalizing $\boldsymbol{\theta}$ with $\mathcal{L}_{\mathrm{CPSD}}$ may not suffice to bootstrap the model from a random initial value.
Against this issue, we assume that roughly half of the songs in each mini-batch of $N$ songs $\mathbf{X}=(\boldsymbol{x}_n)_{n=0}^{N-1}$ are major, the other half being minor.
We denote the corresponding batches of pitch transposition parameters by $\mathbf{C}=(\mathbf{C}[n])_{n}$ and $\mathbf{K}=(\mathbf{K}[n])_{n}$.
We use $T_{\mathbf{C}}\mathbf{X}$ as a shorthand for $((T_{\mathbf{C}[n]} \mathbf{X}_{n,\mathrm{A}}, T_{\mathbf{C}[n]} \mathbf{X}_{n,\mathrm{B}}))_{n}$.
We compute the batch-wise average of mode predictions as
\begin{equation}
\mu_{\boldsymbol{\theta}}^{\mathrm{avg}}(T_{\mathbf{C}}\mathbf{X}) = \dfrac{1}{N} \sum_{n=0}^{N-1}\sum_{L \in \{A,B\}} \boldsymbol{\mu}_{\boldsymbol{\theta}}(T_{\mathbf{C}[n] }\mathbf{X}_{n,\mathrm{L}})[0] 
\end{equation}

and derive the loss function: $\mathcal{L}_{\mathrm{avg}}(\boldsymbol{\theta} \vert\ \boldsymbol{x}, \mathbf{C}) = (\mu_{\boldsymbol{\theta}}^{\mathrm{avg}}(T_{\mathbf{C}}\mathbf{X}) - \frac{1}{2})^2$.

\subsection{Self-supervised learning of major and minor keys (\ourmethod{})}
\label{sub:all-together}
Summing all three terms yields the training loss for \ourmethod{}:
\begin{align}
    \mathcal{L}(\boldsymbol{\theta} \vert\ \mathbf{X}, \mathbf{C}, \mathbf{K}) &= 
    \sum_{n=0}^{N-1} \mathcal{L}_{\mathrm{CPSD}}(\boldsymbol{\theta}\,\vert\,\mathbf{X}_n,\mathbf{C}[n],\mathbf{K}[n])
    \nonumber \\
    &+ \lambda_{\mathrm{S-KEY}} \sum_{n=0}^{N-1}  \mathcal{L}_{\mathrm{S-KEY}}(\boldsymbol{\theta}\,\vert\,\mathbf{X}_n,\mathbf{C}[n],\mathbf{K}[n]) \nonumber \\
    &+ \lambda_{\mathrm{avg}} \mathcal{L}_{\mathrm{avg}}(\boldsymbol{\theta}\,\vert\,\mathbf{X},\mathbf{C}).
    \label{eq:L}
\end{align}
We set the hyperparameters $\lambda_{\mathrm{BCE}}$ and $\lambda_{\mathrm{avg}}$ so that all three terms in the loss $\mathcal{L}$ are of the same order of magnitude at the initialization: $\lambda_{\mathrm{BCE}}=1.5$ and $\lambda_{\mathrm{avg}}=15$.

\section{Application}

\subsection{Training}
\label{sub:training}

STONE was trained on a corpus of 60k songs from the Deezer catalog.
To offer a fair comparison, we begin by training \ourmethod{} on the exact same dataset: see \ref{sub:ssl-sota} and Table \ref{tab:small-scale}.
Later on, we scale up SSL training to 1M songs from Deezer: see \ref{sec:1m} and Table \ref{tab:large-scale}.

We set the duration of segments A and B to 15 seconds. We randomize $c$ uniformly between 0 and 15 semitones, $k$ uniformly between -12 to 12 semitones and $0 \leq k+c \leq 15$.
We train \ourmethod{} for 50 epochs and use a batch size of 128 on the 60k-song corpus versus 100 epochs and a batch size of 256 on the 1M-song corpus.
We use the AdamW optimizer with a learning rate of 0.001 and a cosine learning rate schedule preceded by a linear warm-up.

\subsection{Calibration on C major and A minor scales}
\label{sub:calibration}
The necessity of calibrating two channels separately arises because the model sometimes reaches a local minimum where a shift of fifths exists between the two channels (e.g., C major has the same index as E minor, and as note C in CQT). In this local minimum, $\mathcal{L}_{\mathrm{CPSD}}$ remains low, given that the fifths of a key are considered to be the closest among all keys except for the correct one. $\boldsymbol{\nu}$ would serve as a slightly less accurate pseudo-label than when the model is in its global minimum, however remains a relevant pseudo-label, as demonstrated by empirical results.


We create two synthetic samples, one in C major and another in A minor to calibrate two channels separately. This calibration step is similar to STONE \cite{kong2024stone} except that it operates on a structured output with two modes.

\subsection{Self-supervised and supervised competitors}
\label{subsec:ssl methods}
We compare S-KEY against three self-supervised systems:
\begin{itemize}
\item \textbf{Krumhansl} \cite{krumhansl2001cognitive}. A template matching algorithm for CQT features in which major and minor templates are derived from psychoacoustic judgments, with no machine learning.
\item \textbf{24-STONE} \cite{kong2024stone}. The self-supervised SOTA. It relies on CPSD for equivariance to key signature and on BCE for invariance to mode, with no pseudo-labels.
\item \textbf{$\nu$-STONE}. A simple new method which is an ad hoc procedure using a pre-trained STONE model \cite{kong2024stone}'s prediction of key signature and the rule-based heuristic $\nu$ (Section \ref{sub:pseudo}) for mode prediction which requires no further training. 
\end{itemize}
In addition, we compare S-KEY against the supervised SOTA:
\begin{itemize}
\item \textbf{madmom} \cite{korzeniowski2018genre}. An all-convolutional neural network, trained on a varied corpus (electronic dance music, pop/rock, and classical music) and made available as part of the madmom open-source software library for MIR \cite[v0.16.1]{bock2016madmom}.
\end{itemize}

\subsection{Evaluation datasets and metrics}
\label{sub:evaluation}
We evaluate all systems on the following four datasets, which are labeled according to a taxonomy of 24 major and minor keys:

\begin{itemize}
    \item \textbf{FMAKv2} \cite{kong2024stone}. A derivative of FMAK \cite{wong2023fmak,kong2024stone} which contains 5,489 songs from the Free Music Archive (FMA)\cite{defferrard2017fma}, spread across 17 genres. 
    \item \textbf{GTZAN} \cite{GTZAN_Lerch}. 837 songs from 9 genres. Only songs with a unique key are annotated, therefore no classical music is included.
    \item \textbf{GiantSteps} \cite{giantstep}. 604 two-minute excerpts of electronic dance music (EDM) from commercial songs.
    \item \textbf{SWD} \cite{weiss2021schubert}. 48 classical music pieces composed by Schubert. We only use the first 30s given that key modulations are common in classical music.
\end{itemize}
The MIREX score, as implemented in mir\_eval, is weighted according to the tonal proximity between reference and prediction \cite{raffel2014mir_eval}.
Key signature estimation accuracy (KSEA) assigns a full point to the prediction if it matches the reference and a half point if the prediction is one perfect fifth above or below the reference, and zero otherwise \cite{kong2024stone}.
Mode accuracy assigns a full point if reference and prediction share the same mode (major or minor) and zero otherwise.

\section{Results\protect\footnote{The full training and inference code, along with full details of MIREX score can be found at \href{https://github.com/deezer/s-key}{https://github.com/deezer/s-key}.}}
\label{sec:results}

\subsection{Self-supervised learning from 60k songs}
\label{sub:ssl-sota}

We train all SSL methods on the same 60k-song corpus (see Section \ref{sub:training}) and compare them against a template matching algorithm (Krumhansl \cite{krumhansl2001cognitive}) and the supervised SOTA \cite{korzeniowski2018genre}.

Table \ref{tab:small-scale} summarizes our results on FMAKv2.
\ourmethod{} outperforms the SSL SOTA (24-STONE) as well as Krumhansl's template matching algorithm.
Furthermore, on all three metrics, the performance of \ourmethod{} is within one percentage point of the supervised SOTA.
Thus, \ourmethod{} offers the first proof of feasibility for the value of SSL in full-fledged tonality estimation, i.e., with a taxonomy of 24 keys.
\begin{table}[h!]
    \centering
    \begin{tabular}{l r r r}
         & MIREX (\%) & KSEA (\%) & mode acc. (\%)\\ \hline
        Krumhansl \cite{krumhansl2001cognitive} & 53.4 & 60.1 & 64.9 \\ 
        24-STONE \cite{kong2024stone} & 57.9 & 78.0 & 62.2 \\
        $\nu$-STONE & 67.8 & 79.1 & 74.1 \\
        \textbf{\ourmethod (60k)} & \textbf{72.1}& \textbf{80.3} & \textbf{79.0} \\ \hline
        madmom \cite{korzeniowski2018genre} & 73.1 & 81.3 & 79.3 \\
    \end{tabular}
    \caption{Classification of major and minor keys in the FMAKv2 dataset according to three metrics: \normalfont{MIREX score, key signature estimation accuracy (KSEA) and mode accuracy. Krumhansl's method involves no training, while 24-STONE, $\nu$-STONE, and \ourmethod{} are self-supervised on the same dataset of 60k songs. We include the results of the madmom library as supervised state-of-the-art for reference.}}
    \label{tab:small-scale}
\end{table}

Breaking down the MIREX score into finer-grained  metrics, we observe that the gap in performance between 24-STONE and $\nu$-STONE is primarily attributable to a higher mode accuracy (62.2\% versus 74.1\%) rather than to a higher key signature estimation accuracy (KSEA, 78.0\% versus 79.1\%).
This observation confirms that the rule-based procedure $\boldsymbol{\nu}$ (see Section \ref{sub:pseudo}) is more effective for distinguishing a major key from its relative minor than the BCE-based loss initially developed for 24-STONE.

Unlike $\nu$-STONE, \ourmethod{} is trained from scratch to minimize a joint SSL objective (Equation \eqref{eq:L}) in which $\boldsymbol{\nu}$ plays the role of a pseudo-labeling function.
We posit that this joint optimization creates a virtuous circle: a lower value of the loss improves the informativeness of pseudo-labels, thus making the pretext task less ambiguous, and so forth.
Hence, the data-driven component in \ourmethod{} is able to refine and surpass the ad hoc procedure in $\nu$-STONE.

From $\nu$-STONE to \ourmethod{}, there is not only an improvement in terms of mode accuracy (74.1\% versus 79.0\%), but also in terms of KSEA (79.1\% versus 80.3\%).
This seems to be a benefit of weight sharing and structured prediction in \ourmethod{}.

\subsection{Scaling up to 1M songs}
\label{sec:1m}
Inspired by recent works on large-scale SSL for MIR \cite{li2024mert,meseguer2024experimental}, we retrain \ourmethod{} on a corpus of 1M songs from the Deezer catalog.
Then, we evaluate both versions of \ourmethod{} on FMAKv2 as well as three other annotated datasets: see Section \ref{sub:evaluation}.
Table \ref{tab:large-scale} summarizes our findings.
After SSL on 1M songs, \ourmethod{} performs on-par with the supervised SOTA across all datasets.
Scaling up the training set of \ourmethod{} appears beneficial for three datasets out of four.

\begin{table}[h!]
    \centering
    \begin{tabular}{l | r r r r}
         Dataset & FMAKv2 & GTZAN & GiantSteps & SWD  \\ 
         \#songs & 5,489 & 837 & 604 & 40\\
         \hline
        \ourmethod{} (60k) & 72.1 & 70.9 & 71.7 & 89.0 \\
        \textbf{\ourmethod{} (1M)} & \bf{73.2} & \bf{74.4} & \textbf{72.1} & \bf{90.4} \\
        madmom \cite{korzeniowski2018genre}& 73.1  & 67.9 & 71.0 & 87.7 \\
    \end{tabular}
    \caption{MIREX score (\%) of \ourmethod{} after self-supervised training on 60k or 1M songs. \normalfont{We compare with the madmom package as supervised state of the art. Note: for madmom, we report a score on GiantSteps that is lower than the one reported in the original paper \cite{korzeniowski2018genre}, i.e., 74.6\%, which might due to the different implementations used in madmom and in original paper.}}
    \label{tab:large-scale}
\end{table}

\subsection{Error analysis across genres}

\begin{figure}
    \centering
    \includegraphics[width=0.8\linewidth]{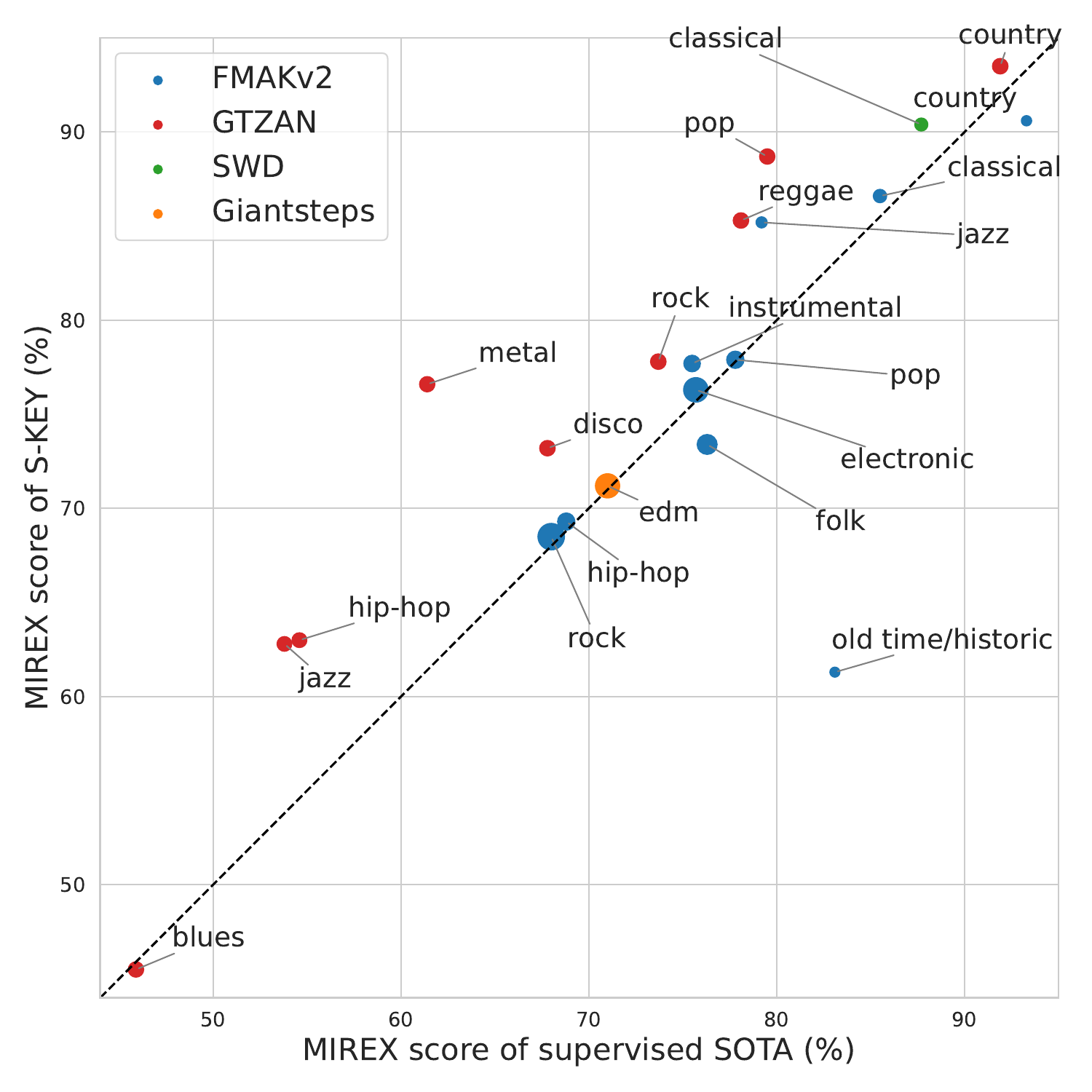}
    \caption{Comparison between the supervised state of the art (x-axis) and \ourmethod{} after self-supervised training on 1M songs (y-axis) in terms of MIREX score, across datasets and genres. The size of each marker is proportional to the number of songs in the corresponding subcorpus. 
    }
    \label{fig:enter-label}
\end{figure}

Figure \ref{fig:enter-label} compares \ourmethod{} versus the supervised SOTA across multiple datasets and genres.
Within GTZAN, both methods achieve a MIREX score above 90\% on \emph{country} and below 50\% on \emph{blues}.
In other words, the gap in MIREX score across genres is much greater than the gap between the two methods over GTZAN as a whole.
Arguably, the MIREX taxonomy of 24 keys is inadequate for blues \cite{jaffe2011something,bluestonality}---likewise, to some extent, for jazz and hip-hop. We leave this important question to future work.

Moreover, the performance for \textit{jazz} shows a large difference between FMAKv2 and Giantsteps. This might be due to the differing genre taxonomies and varying definitions of keys used by annotators\cite{sturm2013gtzan}.

With this caveat in mind, we observe that \ourmethod{} outperforms the supervised SOTA on genres with diverse musical features: e.g., metal, jazz, and reggae.
This suggests that SSL with \ourmethod{} learns invariant representations of tonality.
The only large downgrade from madmom to \ourmethod{} is \emph{old-time/historic}, a small subcorpus of 16 songs in FMAKv2. The small amount of data could lead to a noisy MIREX score.

\subsection{Visualization of \ourmethod{} embeddings}

\begin{figure}[!h]
\includegraphics[width=0.5\linewidth]{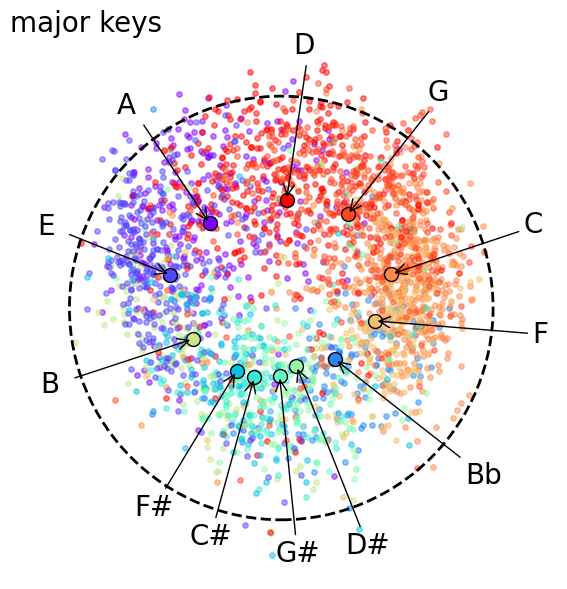}
\includegraphics[width=0.5\linewidth]{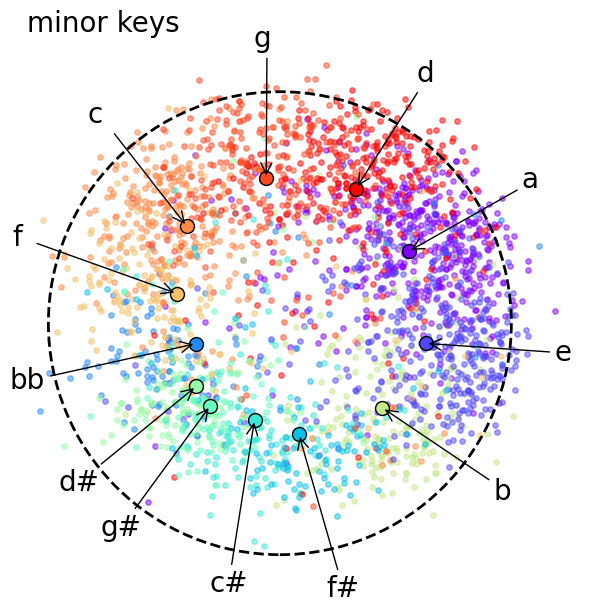}
\caption{2-D visualization of FMAKv2 songs in major and minor keys after self-supervised embedding with S-KEY (trained on 1M songs) and principal component analysis (PCA). Hue indicates key on the circle of fifths, with key labels point at class centroids.}
\label{fig:pca}
\end{figure}

We interpret \ourmethod{} via principal component analysis (PCA) of intermediate features after uniform averaging over time and across ChromaNet channels.
As shown in Figure \ref{fig:pca}, songs in FMAKv2 form a ring pattern which is well explained by the circular progression of fifths, both for major keys (left) and minor keys (right).
Crucially, PCA on CQT features does not show such interpretable patterns.

The circularity of key signatures in \ourmethod{} embeddings results from equivariance in our pretext task design.
This observation is reminiscent of foundational work on self-organizing maps for music cognition \cite{krumhansl2001tonal} and more recent work on unsupervised learning of octave equivalence \cite{lostanlen2020learning}.
Meanwhile, the originality of our finding is that it was obtained by analyzing an unlabled corpus of 1M songs, as opposed to subjective ratings \cite{krumhansl2001tonal} or monophonic sounds \cite{lostanlen2020learning}.
\section{Conclusion}
The promise of self-supervised learning (SSL) in music information retrieval is to harness large unlabeled music corpora to train deep neural networks with little or no annotation effort.
In this article, we have presented \ourmethod{}, an architecture and pretext task for self-supervised learning of 24 keys from audio.
After SSL on 1M songs, \ourmethod{} matches the supervised SOTA on four datasets.
The main limitation behind \ourmethod{} is that its structured prediction is limited to 24 major and minor keys, making it inadequate for certain genres.
Still, the methodological contributions of \ourmethod{}---namely, cross-power spectral density and pitch-invariant pseudo-labeling---could, in principle, apply to blues harmony and modal harmony, given appropriate training data and music-theoretical knowledge.



\bibliographystyle{IEEEbib}
\bibliography{kong2025icassp}

\end{document}